\documentclass{osa-article}
\journal{oe}
\articletype{Research Article}

\usepackage{lineno}
\linenumbers
\newcommand{\br}{{\bf r}}
\newcommand{\bR}{{\bf R}_j}
\newcommand{\bq}{{\bf q}}
\newcommand{\imagunit}{\mathrm{i}}

\newlength{\singlepic}
\newlength{\doublepic}
\begin{document}

\title{Ptychographic reconstruction with wavefront initialization}

\author{Felix Wittwer\authormark{1,2,3} and Peter Modregger\authormark{1,2,*}}

\address{
\authormark{1}Physics Department, University of Siegen, 57072 Siegen, Germany\\
\authormark{2}Center for X-ray and Nano Science CXNS, Deutsches Elektronen-Synchrotron DESY, 22607 Hamburg, Germany\\
\authormark{3}Current address: NERSC, Lawrence Berkeley National Laboratory, Berkeley, California 94720, USA
}

\email{\authormark{*}peter.modregger@uni-siegen.de}

\begin{abstract}
X-ray ptychography is a cutting edge imaging technique providing ultra-high spatial resolutions. In ptychography, phase retrieval, i.e., the recovery of a complex valued signal from intensity-only measurements, is enabled by exploiting a redundancy of information contained in diffraction patterns measured with overlapping illuminations. For samples that are considerably larger than the probe we show that during the iteration the bulk information has to propagate from the sample edges to the center. This constitutes an inherent limitation of reconstruction speed for algorithms that use a flat initialization. Here, we experimentally demonstrate that a considerable improvement of computational speed can be achieved by utilizing a low resolution sample wavefront retrieved from measured diffraction patterns as initialization. In addition, we show that this approach avoids phase singularity artifacts due to strong phase gradients. Wavefront initialization is computationally fast and compatible with non-bulky samples. Therefore, the presented approach is readily adaptable with established ptychographic reconstruction algorithms implying a wide spread use.
\end{abstract}

\section*{Introduction}

X-ray ptychography can be regarded as a combination of scanning X-ray transmission microscopy (STXM) and coherent diffraction imaging (CDI). STXM utilizes a lateral scan of the sample through a focused X-ray beam, which provides the transmission function of the specimen~\cite{Sakdinawat2010}. However, spatial resolution is limited by the focus size. CDI, on the other hand, employs an extended X-ray beam larger than the sample and exploits effective oversampling contained in diffraction patterns with additional constrains during algorithmic retrieval~\cite{Chapman2010}. While spatial resolution provided by CDI is at least in principle wavelength limited, the availability of high quality, large X-ray beams practically limit the sample size.

X-ray ptychography integrates the benefits of STXM and CDI by scanning the specimen through a focal spot with overlapping illuminations, which realizes the oversampling required for stable phase retrieval~\cite{Faulkner2004,Thibault2008,Pfeiffer2018}. This approach allows for samples that are larger than the focal spot while the achieved spatial resolutions are smaller than the focus size. Data analysis is performed by ptychographic reconstruction algorithms, these are able to retrieve the complex wave field of the sample as well as of the illuminating beam, usually referred to as the probe. Therefore, the quality of ptychographic reconstructions is nearly independent from the quality of the X-ray optics rendering this a lensless technique.

Evidently, the success of ptychographic experiments depend on reliable phase retrieval~\cite{Shechtman2014} and, thus, it does not come as a surprise that a wide variety of algorithms have been published. Examples include the extended ptychographic engine (ePIE)~\cite{Maiden2009a}, 3PIE~\cite{Maiden2012}, maximum likelihood refinement~\cite{Thibault2012}, scaled gradient ptychography~\cite{Godard2012}, multi-modal ptychography~\cite{Thibault2013a} or momentum accelerated ptychography~\cite{Maiden2017}. These algorithms tend to reconstruct the complex wave field associated with the sample (and the probe), whereat the phase is wrapped to values modulus $2\pi$. Thus, appropriate phase unwrapping procedures~\cite{Guizar-Sicairos2011,Stockmar2015} have to be used in order to combine 2D ptychography with 3D tomography~\cite{Dierolf2010a,Guizar-Sicairos2011,Diaz2012}. Refractive ptychography~\cite{Chowdhury2019,Wittwer2022}, on the other hand, reconstructs a refractive object function directly and, thus, avoids corresponding phase wrapping artifacts.

% examples of applications: X-ray ptychography has been applied to reveal osteocyte lacunae sites in bone samples~\cite{Dierolf2010a}, Spatial distribution of different components (aluminium, talc, iron oxide) in paint. c~\cite{Chen2013}, Integrated circuits (here an Intel processor) at around 15 nm resolution, showing a view of the active layer with the finest structures.~\cite{Holler2017}

These iterative algorithms start with initial guesses for the probe as well as the object and refine these guesses during the ptychographic reconstruction. Commonly, the probe can be either directly taken from previous scans or well estimated from experimental parameters. As this is generally not the case for the object, a flat initialization is generally used for the object's complex wave field.

Here, we show that the STXM information inherent to ptychography measurements can be exploited to construct a low-resolution estimate of the object's wave field. This estimate is quick to compute and closely resembles the actual object's wave field. We will demonstrate that using such an estimate as the initial guess is especially beneficial for bulky samples increasing the speed of ptychographic iteration while avoiding artifacts associated with large phase gradients.

We will first collect elements of the theoretical basis for ptychography and reconstruct the phase signal of a bulky sample. Thereby, we will demonstrate that the bulk information propagates from the sample edges to the center during iterative reconstruction. Then, we will describe a procedure for the retrieval of the sample's complex wave field directly from the measured diffraction patterns. This low resolution representation of the sample will then be used for wavefront initialization in ptychographic reconstruction, which improves reconstruction speed and avoid phase artifacts associated with large phase gradients. Finally, we will show the reconstruction of a worst case sample (i.e., a non-bulky sample with structures smaller than the probe) is not impaired by utilizing wavefront initialization.

\section*{Ptychographic reconstruction}

\begin{figure}[thbp]
    \centering
      \includegraphics[width=.6\linewidth]{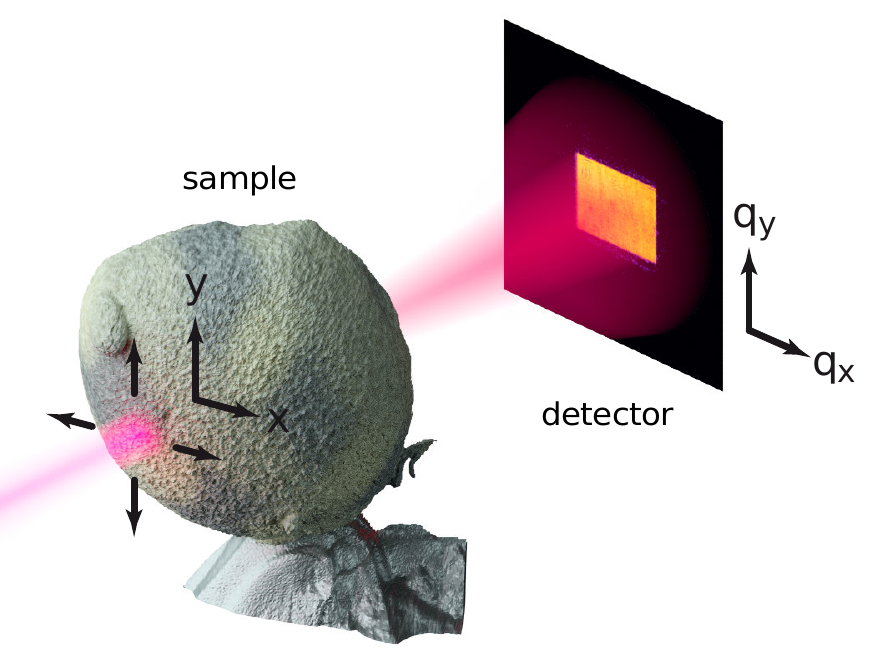}
    \caption{Sketch of the experimental setup for ptychography.}
    \label{fig:ptychography_setup}
\end{figure}

%  [WHY DOES THE BEAM LOOK SQUARE? Ftrans of Ftrans of kb aperature]

In the following, we will re-iterate elements of the frame work for refractive ptychography~\cite{Wittwer2022} for the convenience of the reader. In the X-ray regime the complex refractive index $n$ of a material is commonly expressed as
\begin{equation}
    n=1-\delta+\imagunit\beta,
\end{equation}
with $\delta$, the refractive index decrement and $\beta$, the absorption index. For samples that are sufficiently thin to avoid internal diffraction, the complex wave field $O(\br)$ at the object plane point $\br = (x,y)$ (Fig.~\ref{fig:ptychography_setup}) after transmission is given by
\begin{equation}
    O(\br) = \exp \left(\imagunit\tilde O(\br)\right) = \exp \left( \imagunit k \int\!\!dz\, \left(n(x,y,z)-1\right) \right)
\end{equation}
with $k$, the modulus of the wave vector and $z$, the direction along the optical axis. The goal of refractive ptychography is to retrieve the refractive object function $\tilde O(\br)$. In the experiment the object function is illuminated by a focused X-ray beam with the complex wave field of the probe $P(\br-\bR)$ at $\bR$ the $j$-th scan position. The resulting complex wave field $\Psi (\br)$ is given as
\begin{equation}\label{eq:wavefield}
    \Psi_j (\br) = \exp \left(\imagunit\tilde O(\br)\right) \cdot P(\br-\bR).
\end{equation}
and the observable intensity is
\begin{equation}\label{eq:obs_int}
    \hat I_j(\bq) = \left| \mathcal{F}\left\{ \Psi_j (\br) \right\}\!(\bq)  \right|^2
\end{equation}
with $\mathcal{F}$, the Fourier transform and $\bq = (q_x,q_y)$ the variables conjugate to $(x,y)$. Reconstruction is achieved by minimizing the cost function $L$
\begin{equation}\label{eq:cost_function}
    L = \sum_{j,\bq} \left| \sqrt{\hat I_j(\bq)}  - \sqrt{\hat D_j(\bq)}  \right|^2
\end{equation}
with $\hat D_j(\bq)$, the diffraction pattern measured at scan position $\bR$. The use of square roots takes into account Poisson statistics of photon shot noise~\cite{Godard2012}. Minimization of $L$ is usually performed iteratively for example by the statistical gradient descent scheme~\cite{Maiden2009a,Wittwer2022}. Here, the object wave field $O_n(\br)$ and the probe wave field $P_n(\br))$ are updated in each iteration step $n$ as a loop over all diffraction measurements $\hat M_j(\bq)$ in a random order. With respect to each diffraction measurement the update is performed by first calculating the current wave field at the detector
\begin{equation}
    \hat \psi_j (\bq) = \mathcal{F}\left\{ \exp \left(\imagunit \tilde O_n(\br)\right) \cdot P_n(\br-\bR) \right\}.
\end{equation}
Then the modeled amplitude $| \hat \psi_j (\bq)|$ is replaced with the amplitude of the measurements $\sqrt{\hat D_j(\bq)}$ and the resulting wave field is propagated back to the object plane
\begin{equation}
    \psi_j' (\br) = \mathcal{F}^{-1} \left\{ \sqrt{\hat D_j (\bq)} \cdot \frac{\hat \psi_j (\bq)}{| \hat \psi_j (\bq)|}    \right\}
\end{equation}
with $\mathcal{F}^{-1}$, the inverse Fourier transform. Finally, the object wave field is updated by
\begin{equation}
    \tilde O_{n+1}(\br) \leftarrow \tilde O_n(\br) + \alpha \frac{\left(\imagunit\psi (\br-\bR) \right)^*}{\mathrm{max}|\psi (\br-\bR)|^2}\cdot
    \left( \psi' (\br) - \psi (\br) \right)
\end{equation}
and the probe wave field by
\begin{equation}
    P_{n+1}(\br) \leftarrow P_{n}(\br) + \beta \frac{\exp \left(-\imagunit\tilde O_n^*(\br+\bR) \right)}{\mathrm{max}|\exp \left(-\imagunit\tilde O_n(\br+\bR) \right)|^2}\cdot
    \left( \psi' (\br) - \psi (\br) \right).
\end{equation}
The update strength is tuned by the parameters $\alpha$ and $\beta$. Initialization of the iterative procedure refers to the starting values for the object function $\tilde O(\br)$ and the probe $P(\br)$. Usually, a flat initialization for the object function is chosen, i.e.
\begin{equation}\label{eq:flat_init}
O_0 (\br) = \exp \left(\imagunit\tilde O_0(\br)\right) = 1.
\end{equation}
The complex wave field of the probe $P(\br)$ is usually well characterized, which allows a realistic initialization.  

For the experimental demonstration of our proposed approach we will reuse a previously published ptychographic scan of a micrometeorite~\cite{Wittwer2022}. The essential experimental parameters were as follows. The scan was carried out at the P06 beamline of PETRA III (DESY, Hamburg)~\cite{Schroer2017a,Schropp2020} using the combined Micro- and Nanoprobe setup~\cite{Schropp2019}. The X-ray beam with a photon energy of 18~keV was focused by two orthogonal Kirkpatrick–Baez mirrors to a probe size of 300~nm~$\times$~200~nm at the position of the sample. The micrometeorite with a diameter of about 80~$\mu$m was scanned in fly-mode over a field of view of 100~$\mu$m~$\times$~100~$\mu$m in steps of 200~nm resulting in 250,000 scan points. The diffraction patterns were recorded with an EIGER~500k (Dectris, Switzerland) located 8.75~m downstream of the sample with a dwell time of 1~ms. The central 128~$\times$~128 pixels were used for ptychographic reconstruction yielding a effective pixel size of 62.8~nm. Iterative reconstruction was performed based on refractive ptychography as described above and by scaled gradient descent~\cite{Godard2012} with an additional momentum accelerated update~\cite{Maiden2017} every second iteration step.

\begin{figure}[htbp]
    \centering
    \begin{tabular}{cc}
      \includegraphics[width=\doublepic]{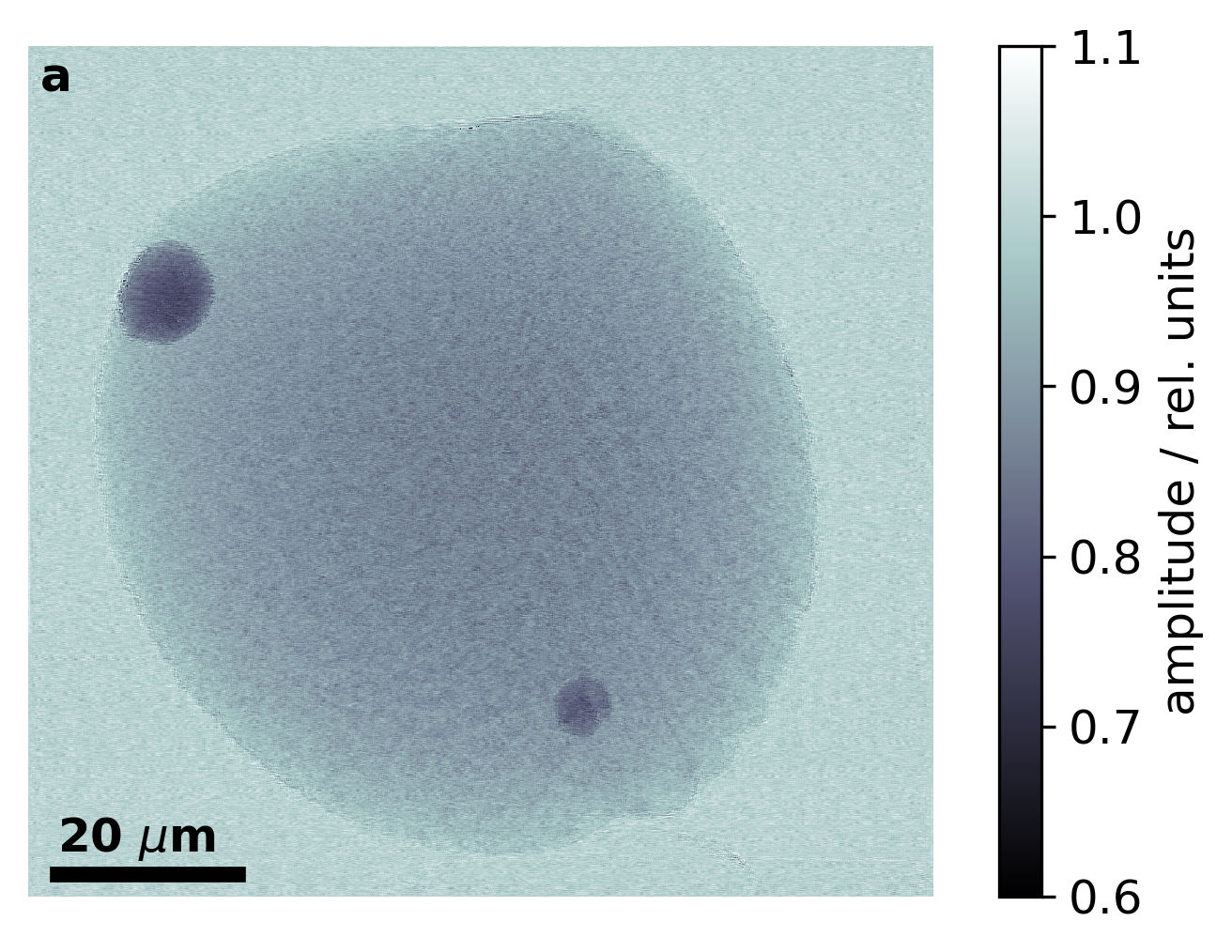}   &
      \includegraphics[width=\doublepic]{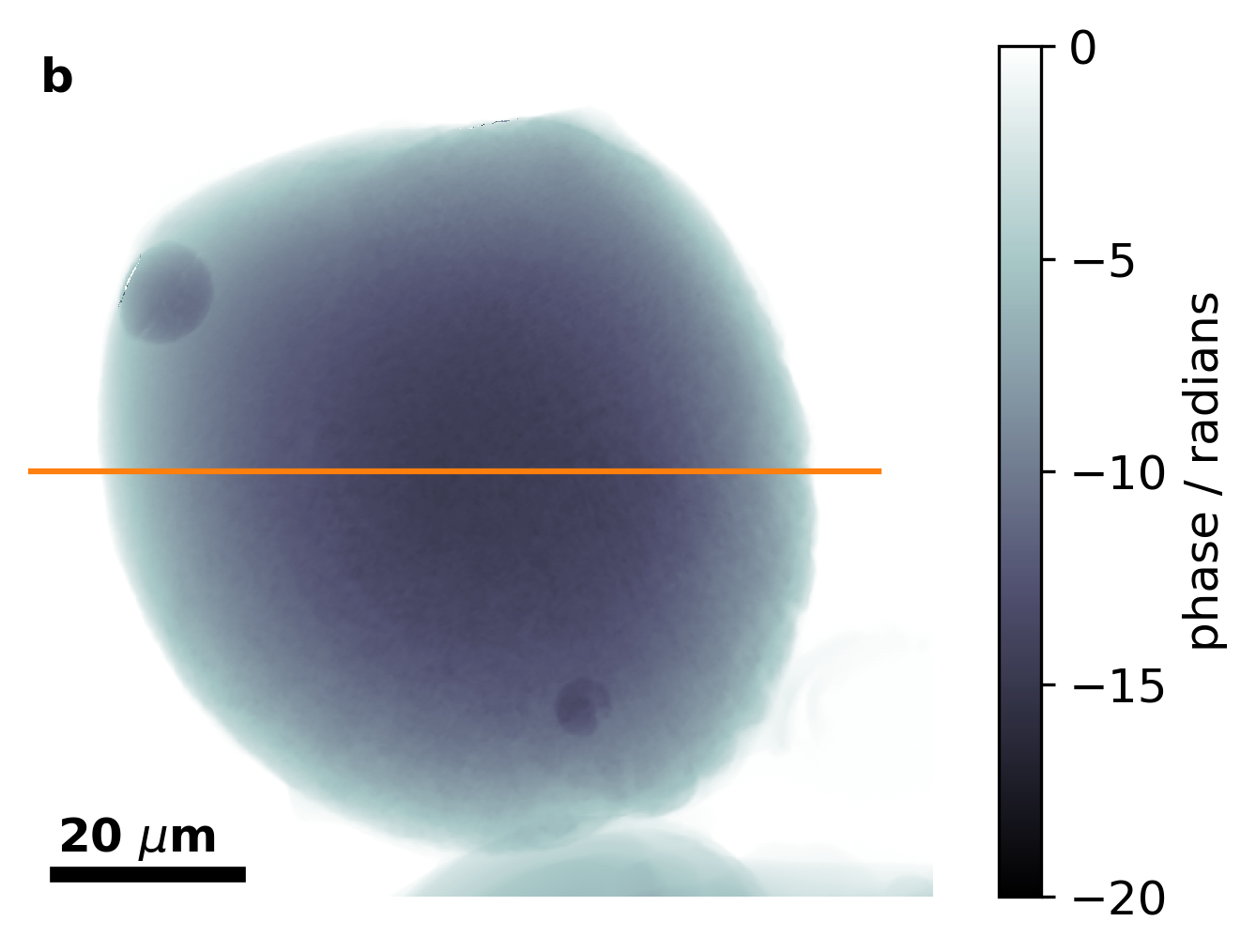}  
    \end{tabular}
    \caption{Modulus (a) and phase (b) of the micrometeorite as retrieved by refractive ptychography. The field of view is 100~$\mu$m~$\times$~100~$\mu$m. The horizontal line in (b) indicates the position of the lines profiles in Fig.~\ref{fig:propagation_edge}.}
    \label{fig:standard_ptycho_result}
\end{figure}

Figure~\ref{fig:standard_ptycho_result} shows the result of refractive ptychographic reconstruction with flat initialization (eq.~\ref{eq:flat_init}). Artifacts at the left and the top edges of the sample appear in the reconstruction of the phase, which can be attributed to large phase gradients (see below). Clearly, the micrometeorite constitutes a bulky sample, which is about 400 times larger than the probe.

\begin{figure}[htbp]
    \centering
    \includegraphics[width=\singlepic]{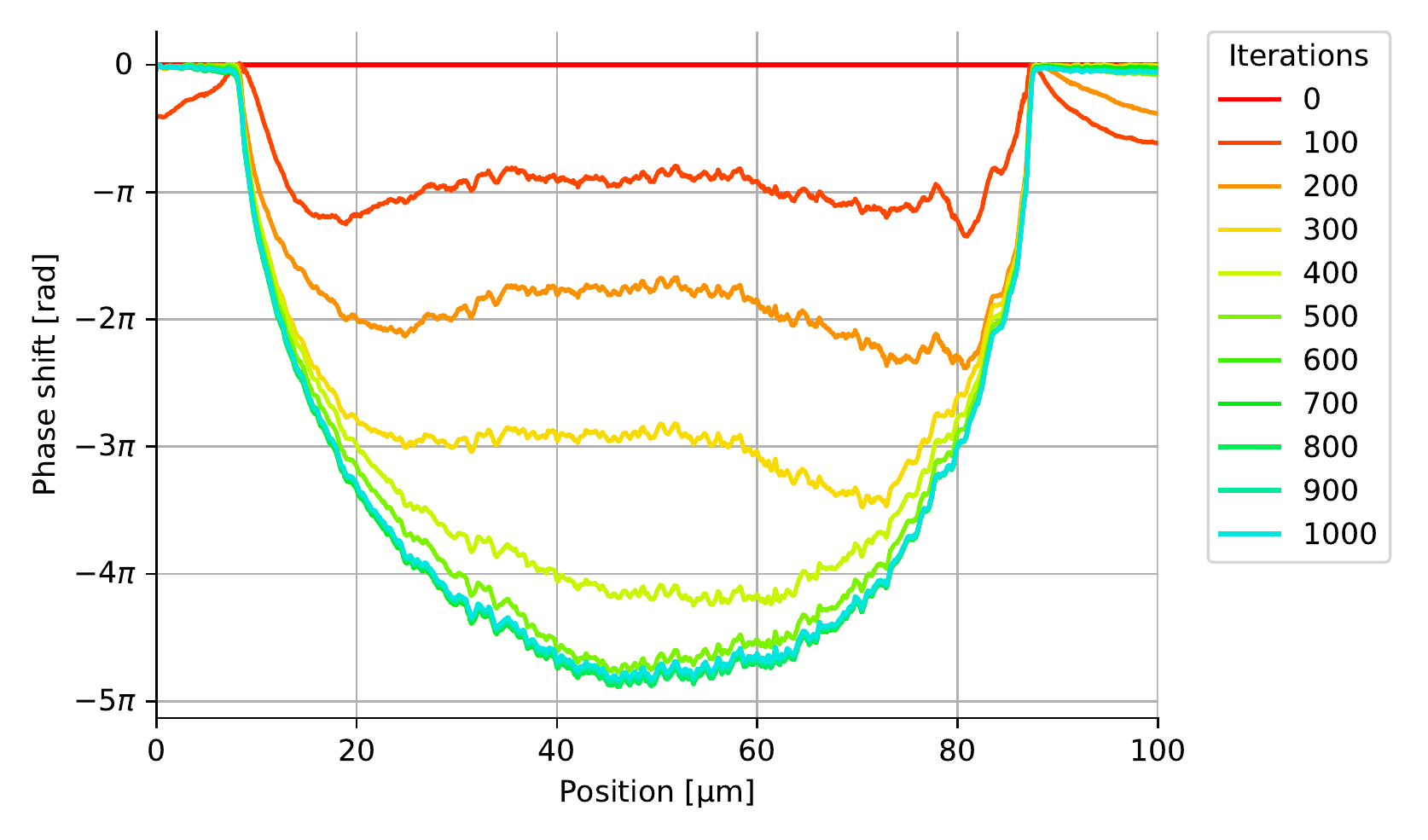}
    \caption{Horizontal line profiles (Fig.~\ref{fig:standard_ptycho_result}b) of the reconstructed phase signal through the middle of the micrometeorite as a function of iteration steps. With each iteration, the reconstructed phase improves from the edges of the sample towards the center.}
    \label{fig:propagation_edge}
\end{figure}

Figure~\ref{fig:propagation_edge} illustrates the behaviour of the reconstructed phase signal as a function of iteration steps for this bulky sample. It is evident that during iteration the bulk information propagates from the edges of the sample to the center, which can be explained as follows. Phase information is predominately encoded in the lateral offset of the diffraction patterns due to refraction or, equivalently, phase gradients. Thus, there is little difference between the exterior of the sample and its center as the phase gradients in both regions are negligible. Therefore, bulk phase information is most prevalent only at the edges of the sample. In each iteration step this information can only be shared between neighboring scan points, which leads to a large number of required iteration steps for bulky samples. 
Initialising the ptychographic reconstruction with a sample wave field that already carries bulk information can improve reconstruction speed considerably in contrast to a flat initialization.

\section*{Moment analysis}

The sample's complex wavefront $O(\bR) = o(\bR) e^{\imagunit\Phi (\bR)}$ with the object's transmission $o(\bR)$ and the object's phase $\Phi(\bR)$ will be constructed from the moments $M_{uv}$ of the measured diffraction patterns $\hat D_j(\bq)$, which are given as~\cite{Thibault2009,Bunk2009b,Modregger2014c}
\begin{equation}
M_{uv}(\bR) = \int\!\! d\bq \, (q_x)^u (q_y)^v \hat D_j(\bq)
\end{equation}
with $u$ and $v$ both integers indicating the horizontal and vertical order of the moment, respectively. In the following, we will be only interested in the three moments up to the first order. The transmission signal of the object corresponds to $M_{00}$ in a straight forward way:
\begin{equation}
    o^2(\bR) = \dfrac{\int\!\! d\bq \, \hat D_j(\bq)}{\int\!\! d\bq \, \hat D_{\mathrm{flat}}(\bq)}
\end{equation}
with $D_{\mathrm{flat}}(\bq)$ a diffraction pattern taken in a region outside the object for the purpose of normalization.

Typically, the horizontal differential phase signal of the object $\Phi_x(\bR) = \partial_x \Phi(\bR)$ is associated with $M_{10}$ and the vertical differential phase signal $\Phi_y(\bR) = \partial_y \Phi(\bR)$ with $M_{01}$. However, this association relies on at least two implicit assumptions as demonstrated in~\cite{Thibault2009} and which we will show in the following. 
It has been analytically demonstrated that the moments of the diffraction patterns $M_{uv}$ are connected to the complex input wavefield in direct space $\Psi(\br)$ by a simple integral~\cite{Modregger2017}. 
For $M_{10}$ this is given as
\begin{equation}
    M_{10}(\bR) = \frac{1}{-iK} \int\!\! d\br \, \Psi_j^*(\br)\, \partial_x \Psi_j(\br)
\end{equation}
with the modulus of the wave vector $K$ and an analogous equation for $M_{01}$, which involves the partial derivative $\partial_y$. 
Using eq.~(\ref{eq:wavefield}) and $P(\br-\bR) = p(\br-\bR) e^{i\xi (\br-\bR)}$ with the probe's transmission $p(\br-\bR)$ and the probe's phase $\zeta(\br-\bR)$ leads to
\begin{equation}\label{eq:2parts}
    M_{10}(\bR) = \int\!\! d\br \, \left[ 
    p^2(\br-\bR) o^2(\br)\, \partial_x \Phi (\br) +  
    p^2(\br-\bR) o^2(\br)\, \partial_x \xi (\br)
    \right],
\end{equation}
where terms involving derivatives of the transmission signals vanish, since the probe has a finite support: $\int \!\! d\br \, p(\br)\, \partial_x p(\br) = p^2(\br)/2|_{-\infty}^{\infty} = 0$. The first term in eq.~(\ref{eq:2parts}) corresponds to the object's phase gradient, which is of interest here. The second term constitutes a contribution of the probe to $M_{10}$ and is non-zero in the combined case of a non-vanishing absorption signal of the object $o(\bR)\neq 1$ and a non-vanishing phase gradient of the probe $\partial_x \xi(\br-\bR) \neq 0$. The latter is typically the case if the object is located outside of the beam focus, the optics are aberrated, or if a purposefully structured probe is used.

% For scan regions outside of the object this term reduces to a standard flat field correction in this context, i.e.: $\int\!\! d\br \, p^2(\br-\bR)\, \partial_x \xi (\br)$.

Assuming that the second term in eq.~(\ref{eq:2parts}) can be neglected, the differential phase signals of the object in the area of illumination defined by the probe $P(\br-\bR)$ at the scan point $\bR$ can be estimated as
\begin{equation}\label{eq:diff_phase_naive1}
    \Phi_{x}(\bR) = \dfrac{\int\!\! d\bq \, q_{x} \hat D_j(\bq)}{\int\!\! d\bq \,  \hat D_j(\bq)} -
                 \dfrac{\int\!\! d\bq \, q_{x} \hat D_{\mathrm{flat}}(\bq)}{\int\!\! d\bq \,  \hat D_{\mathrm{flat}}(\bq)}
\end{equation}
and
\begin{equation}\label{eq:diff_phase_naive2}
    \Phi_{y}(\bR) = \dfrac{\int\!\! d\bq \, q_{y} \hat D_j(\bq)}{\int\!\! d\bq \,  \hat D_j(\bq)} -
                 \dfrac{\int\!\! d\bq \, q_{y} \hat D_{\mathrm{flat}}(\bq)}{\int\!\! d\bq \,  \hat D_{\mathrm{flat}}(\bq)},
\end{equation}
where correction terms account for the pixel position of the flat field beam in the detector. 

If the second term in eq.~(\ref{eq:2parts}) cannot be neglected, the above equations have to be corrected. For this, the influence of the object's transmission and the probe's differential phase signal must be determined.  This can be done by either using the second term directly or -- more conveniently -- by calculating virtual diffraction patterns $\hat V_j(\bq)$ provided by the pure absorption signal of the object according to
\begin{equation}
    \hat V_j (\bq) = \left| \mathcal{F}\left\{o(\bR) P(\br-\bR) \right\}\!(\bq) \right|^2.
\end{equation}
The moments $M_{10}$ and $M_{01}$ of these virtual diffraction pattern correspond exactly to the second term in eq.~(\ref{eq:2parts}). Thus, the differential phase signals can be estimated in this case as
\begin{equation}\label{eq:diff_phase_corr1}
    \Phi_{x}(\bR) = \dfrac{\int\!\! d\bq \, q_{x} \hat D_j(\bq)}{\int\!\! d\bq \,  \hat D_j(\bq)} -
                 \dfrac{\int\!\! d\bq \, q_{x} \hat V_j(\bq)}{\int\!\! d\bq \,   \hat V_j(\bq)}
\end{equation}
and
\begin{equation}\label{eq:diff_phase_corr2}
    \Phi_{y}(\bR) = \dfrac{\int\!\! d\bq \, q_{y} \hat D_j(\bq)}{\int\!\! d\bq \,  \hat D_j(\bq)} -
                 \dfrac{\int\!\! d\bq \, q_{y} \hat V_j(\bq)}{\int\!\! d\bq \, \hat V_j(\bq)}.
\end{equation}

\section*{Wavefront retrieval}

\begin{figure}
    \centering
    \begin{tabular}{cc}
     \includegraphics[width=0.8\doublepic]{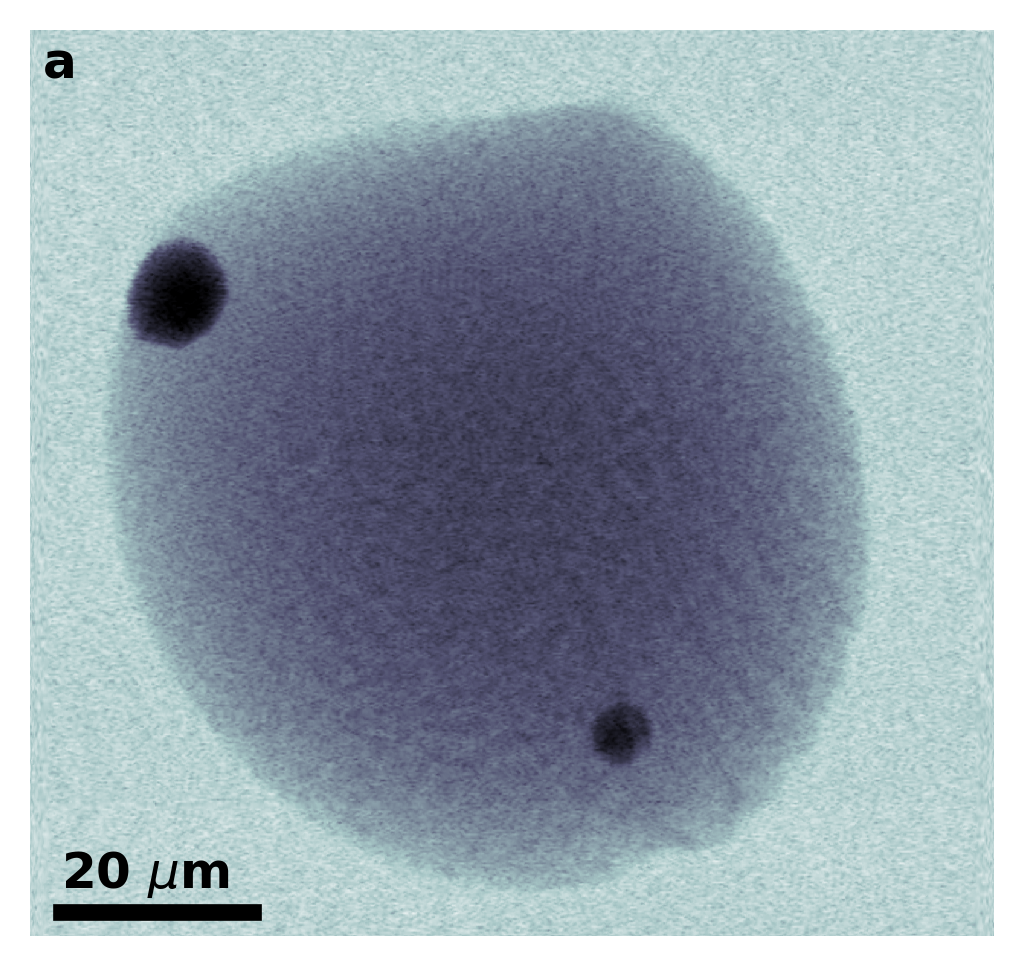} &
     \includegraphics[width=0.8\doublepic]{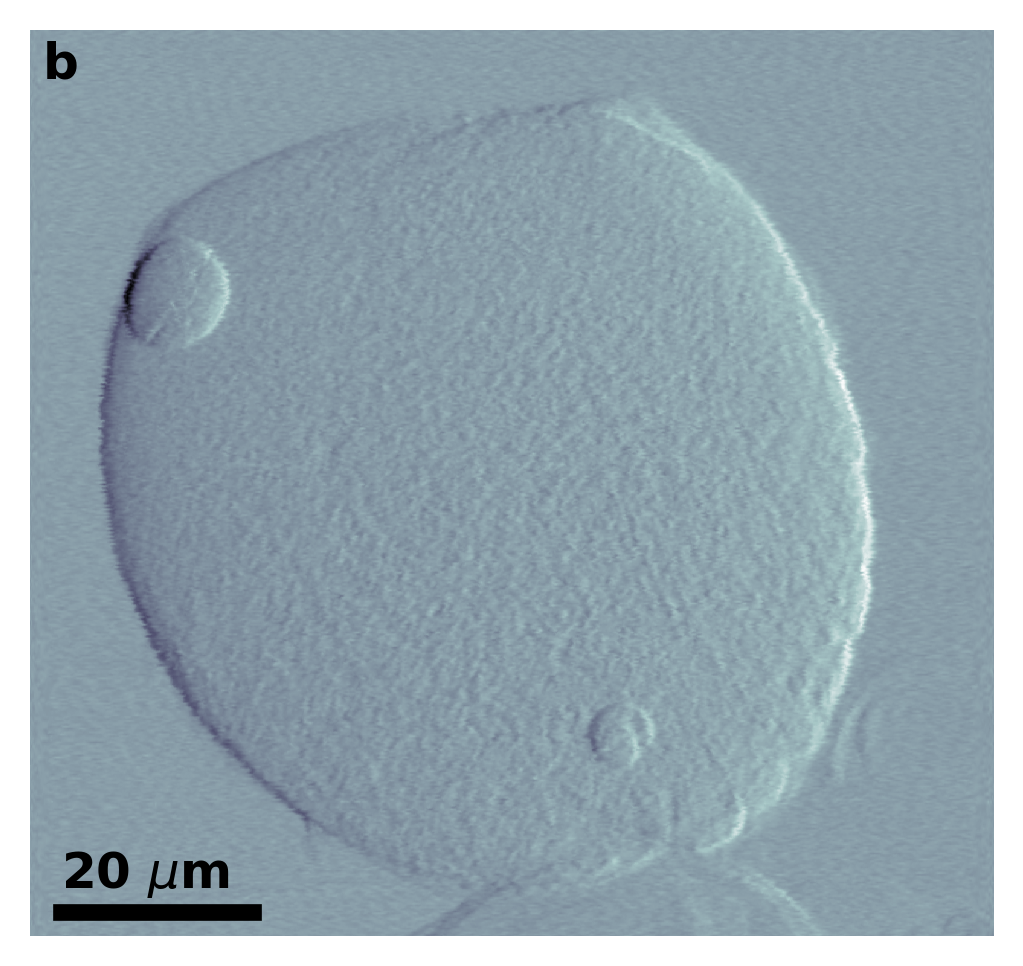} \\
     \includegraphics[width=0.8\doublepic]{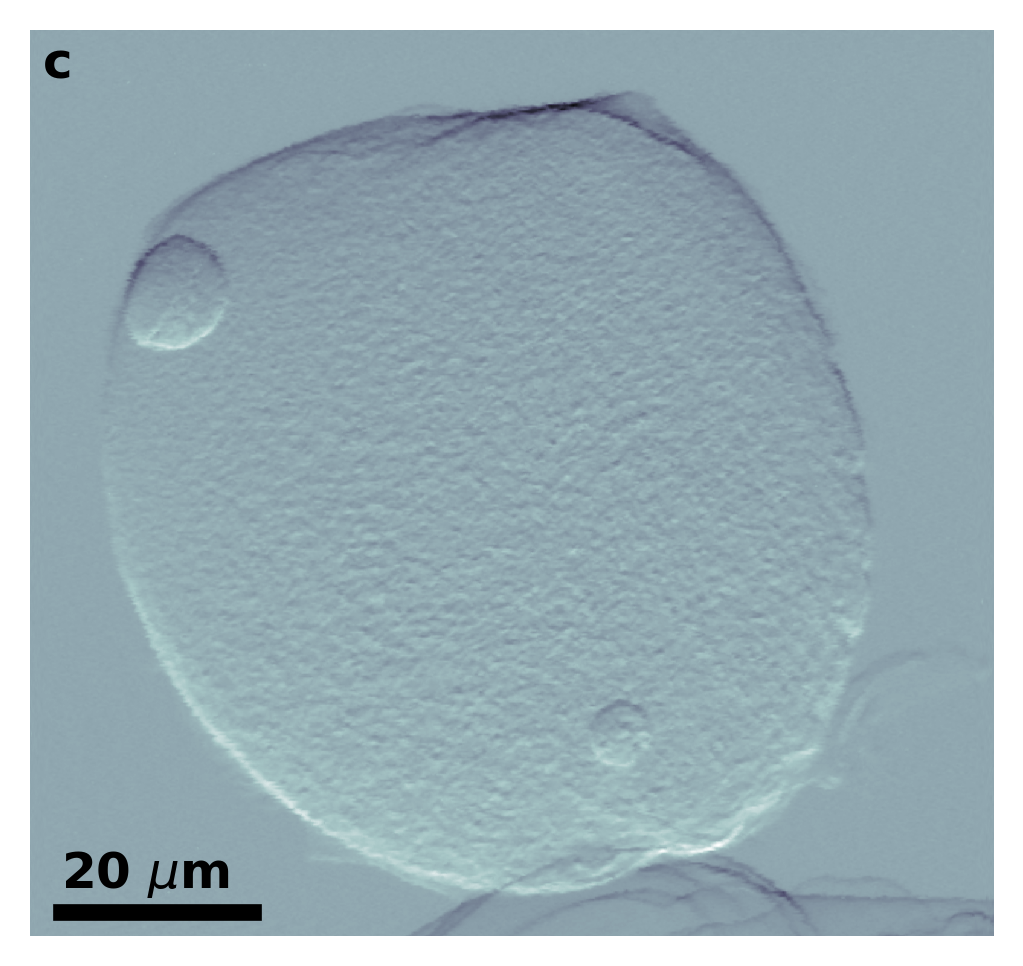} &
     \includegraphics[width=0.8\doublepic]{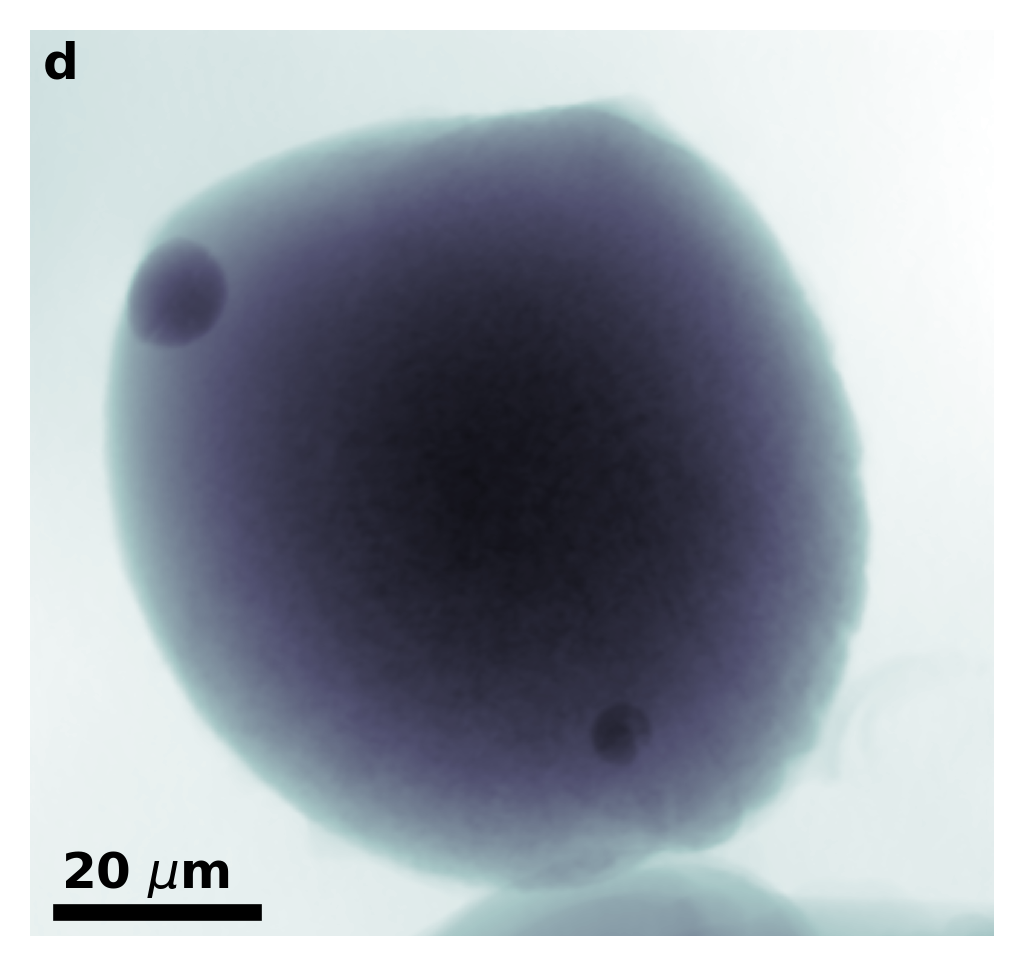}
    \end{tabular}
    \caption{Non-iterative wavefront retrieval of the sample's complex wave field used for subsequent initialization. The moment analysis of diffraction patterns provide the absorption (a), horizontal (b) and vertical phase gradients (c). The combination of the absorption image (a) and the non-iteratively retrieved phase (d) provide a complex wave field of the sample with low spatial resolution.}
    \label{fig:wavefront_retrieval}
\end{figure}

In order to retrieve the phase image $\Phi(\bR)$ from the estimated differential phase images $\Phi_x(\bR)$ and $\Phi_y(\bR)$ we use the non-iterative, boundary-artifact-free wavefront reconstruction presented in~\cite{Bon2012}. This approach starts with constructing an antisymmetric extension of the inputs
\begin{equation}
    \bar \Phi_x = \begin{bmatrix}
                 -\Phi_x(-x,-y) & \Phi_x(x,-y) \\
                 -\Phi_x(-x,y) & \Phi_x(x,y)
                 \end{bmatrix}
\end{equation}
and
\begin{equation}
    \bar \Phi_y = \begin{bmatrix}
                 -\Phi_y(-x,-y) & -\Phi_y(x,-y) \\
                 \Phi_y(-x,y) & \Phi_y(x,y)
                 \end{bmatrix}.
\end{equation}
Then the integrated image is retrieved by calculating
\begin{equation}
    \bar \Phi(\bR) = \mathcal{F}^{-1}\left\{
    \frac{\mathcal{F}\left\{ \bar \Phi_x\right\}\!(\bq) +i  \mathcal{F}\left\{ \bar \Phi_y\right\}\!(\bq) }
    {{q_x}+i{q_y}}
    \right\},
\end{equation}
which was also published in~\cite{Kottler2007c}. Final cropping of $\bar \Phi(\bR)$ to the region of interest yields a low resolution representation of the object phase $\Phi(\bR)$. The results of non-iterative wavefront retrieval for the micrometeorite sample are illustrated in Fig.~\ref{fig:wavefront_retrieval}. Here, the phase gradient of the probe was negligible, so that eqs.~(\ref{eq:diff_phase_naive1}) and~(\ref{eq:diff_phase_naive2}) have been used for the estimation of the phase gradients.

\section*{Ptychographic reconstruction with wavefront retrieval}

With the availability of the low resolution versions of the absorption and the phase image the initialization for the sample's complex wave field is given by
\begin{equation}\label{eq:wavefront_init}
    O_0(\br) = o(\bR) \cdot \exp \left( \imagunit \Phi(\bR) \right),
\end{equation}
where interpolation between the coordinate systems defined by $\br$ and $\bR$ is used as necessary.

\begin{figure}
    \centering
    \begin{tabular}{cc}
        \includegraphics[width=\doublepic]{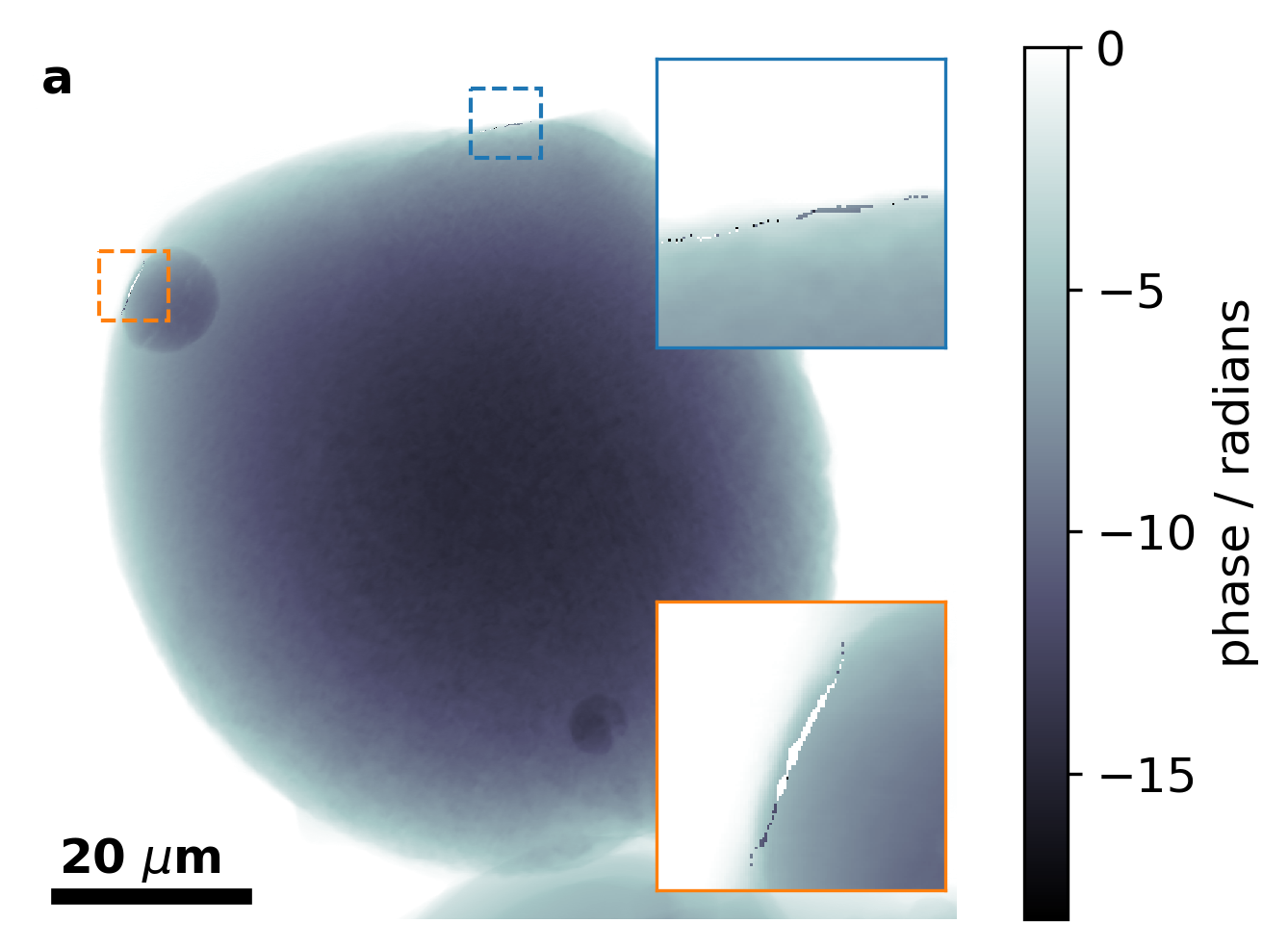}   &
        \includegraphics[width=\doublepic]{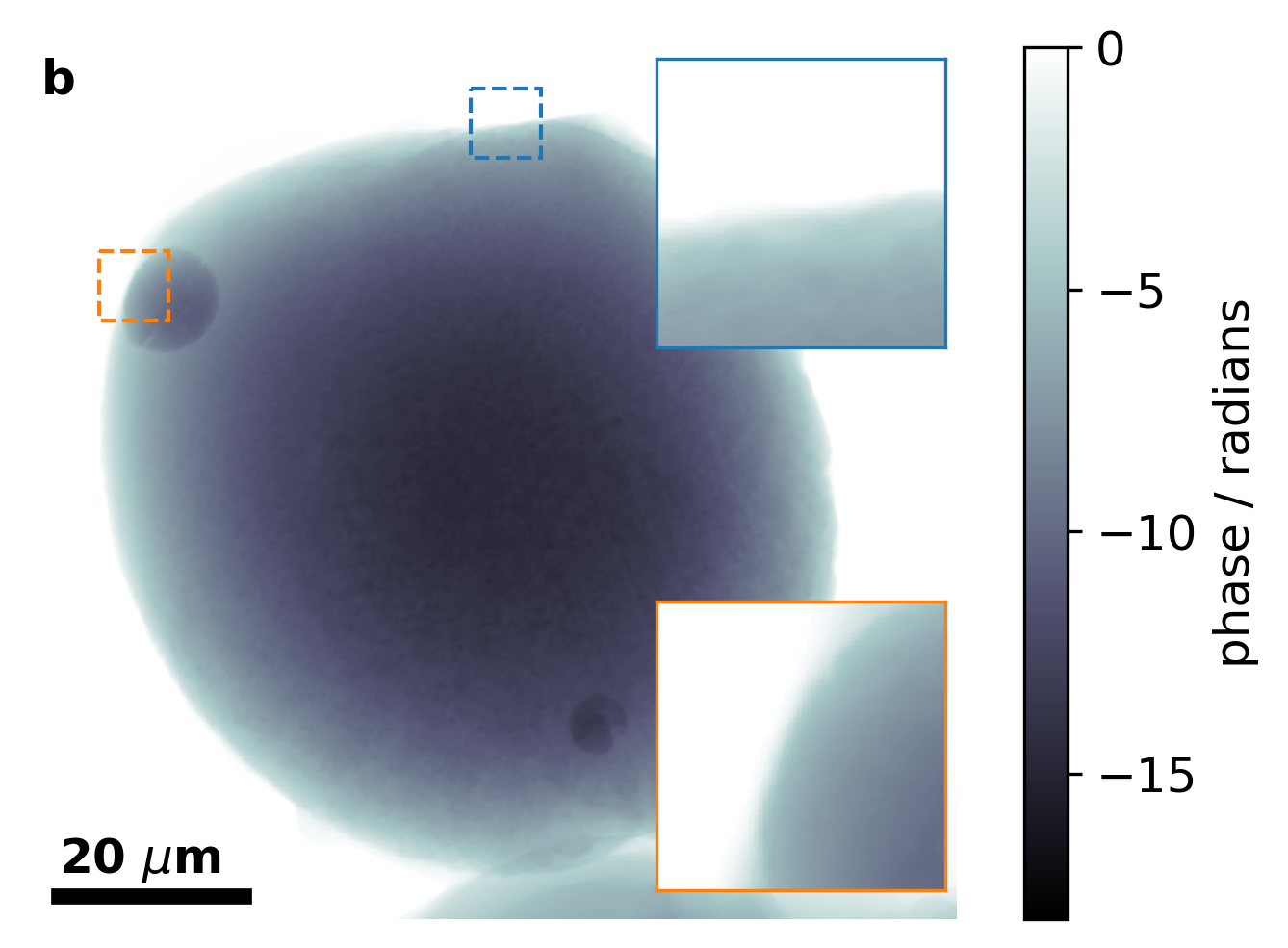}
    \end{tabular}
    \caption{The retrieved phase signals of the micrometeorite with flat initialization (a) and wavefront initialization (b) are broadly similar as expected. The insets show magnified regions, where phase singularities occur with flat initialization and are located at positions of the largest phase gradients (cmp. to Fig.~\ref{fig:wavefront_retrieval} b and c). These artifacts are avoided by using wavefront initialization as shown by the inset in (b).}
    \label{fig:micrometeorite_ptycho_overview}
\end{figure}

%\begin{figure}[htbp]
%    \centering
%    \includegraphics[width=\singlepic]{figures/micrometeorite_ptycho_detail.pdf}
%    \caption{Details of the phase reconstructions shown in %Fig.~\ref{fig:micrometeorite_ptycho_overview}. The flat initialization (a\& c) resulted in some %phase singularity artifacts, which are absent when using wavefront initialization (b \& d). The %scale bar is 5~$\mu$m and the field of view is 18~$\mu$m.}
%    \label{fig:micrometeorite_ptycho_detail}
%\end{figure}

Figure~\ref{fig:micrometeorite_ptycho_overview} shows the resulting phase signal of the micrometeorite after 1000 iteration steps for flat initialization in panel (a) and wavefront initialization in panel (b). As expected the images are fairly similar. However, in the reconstruction with flat initialization (a) phase singularities are present at the left and the top border of the sample coinciding with the locations of the largest phase gradients visible in Fig.~\ref{fig:wavefront_retrieval} (b) and (c). These artifacts are absent for ptychographic reconstruction with wavefront initialization (insets in Fig.~\ref{fig:micrometeorite_ptycho_overview}b).

\begin{figure}
    \centering
    \includegraphics[width=0.95\textwidth]{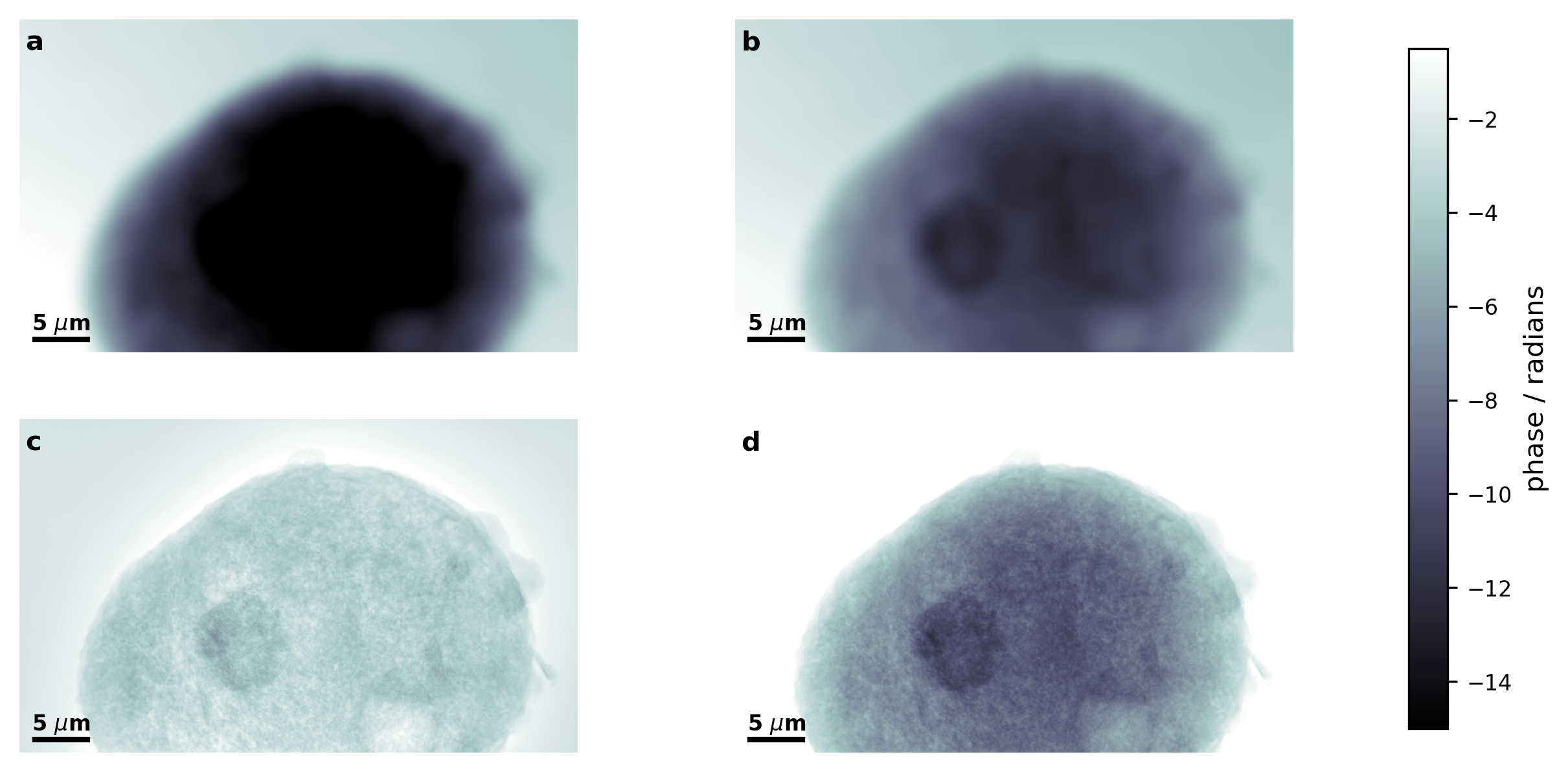}
    \caption{Estimated phase and ptychographic reconstruction of a fluid catalytic cracking catalyst particle. (a) Estimated phase without the necessary correction for the probe influence (i.e., using eqs.~\ref{eq:diff_phase_naive1} and~\ref{eq:diff_phase_naive2}). (b) Estimated phase with the necessary correction for the probe influence (i.e., using eqs.~\ref{eq:diff_phase_corr1} and~\ref{eq:diff_phase_corr2}). (c) Ptychographic reconstruction of the object's phase signal with flat initialization. (d) Ptychographic reconstruction of the object's phase signal with wave front initialization. The data set was taken from~\cite{Odstrcil2019b}.}
    \label{fig:csaxs}
\end{figure}

In order to demonstrate the versatility of the wavefront initialization for ptychographic reconstructions, the proposed approach was further applied to a ptychographic data set which was acquired at the cSAXS beamline of the Swiss Light Source, PSI, Switzerland~\cite{Holler2014}. Here a fluid catalytic cracking catalyst particle with a diameter of 20~$\mu$m was scanned with a photon energy of 6.2~keV over 2344 positions in a spiral trajectory. The average step size was 0.8~$\mu$m covering a field of view of 50~$\mu$ by 30~$\mu$. The pixel size of the reconstruction was 29~nm.More details of the setup and scan procedure can be found in the original work~\cite{Odstrcil2019} and the data set is available online~\cite{Odstrcil2019b}. One of the goals of Odstr{\v{c}}il\,\emph{et al} was to show that purposefully structured probes can improve ptychographic reconstructions. However, the data set used in the following (named "FCC\_particle\_FZP\_11\_dataset\_id1.mat") was acquired without a structured probe, but since the object was placed out of the beam focus, the probe still showed a noticeable phase gradient. 

Figure~\ref{fig:csaxs} summarizes the results of ptychographic reconstruction of this data set in the context at hand. Panel (a) shows the estimation of the object's phase without appropriate correction for the probe's phase gradient (i.e., using eqs.~(\ref{eq:diff_phase_naive1}) and~(\ref{eq:diff_phase_naive2})). Apparently, the object's phase shift was significantly overestimated. Panel (b) demonstrates that the appropriate correction (i.e., using  eqs.~(\ref{eq:diff_phase_corr1}) and~(\ref{eq:diff_phase_corr2})) solves this issue. Panel (c) displays the result of refractive pytchographic reconstruction and flat initialization after 400 iteration steps (equivalent to Fig.~5a in~\cite{Odstrcil2019}). Panel (d) shows the result of ptychographic reconstruction with wavefront initialization, i.e. the phase estimation in panel (b) plus the corresponding absorption signal was used to initialize the iterative minimization. Remarkably, this result is equivalent to Fig.~5d in~\cite{Odstrcil2019} without using a structured probe. This has potential benefits for combining ptychography with other X-ray techniques that would suffer from structured probes, such as X-ray fluorescence.

\begin{figure}
    \centering
    \includegraphics[width=1.1\doublepic]{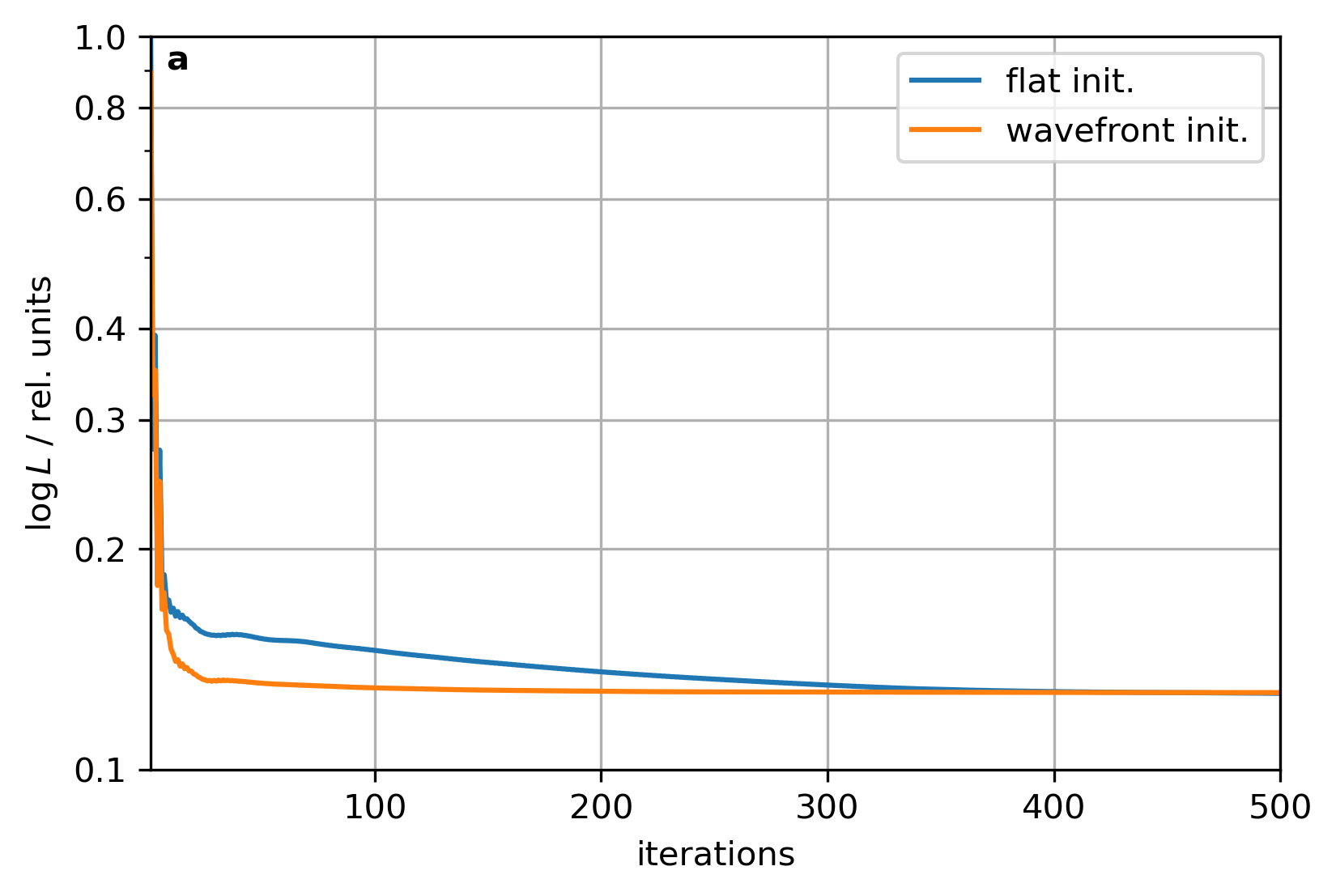}
    \includegraphics[width=1.1\doublepic]{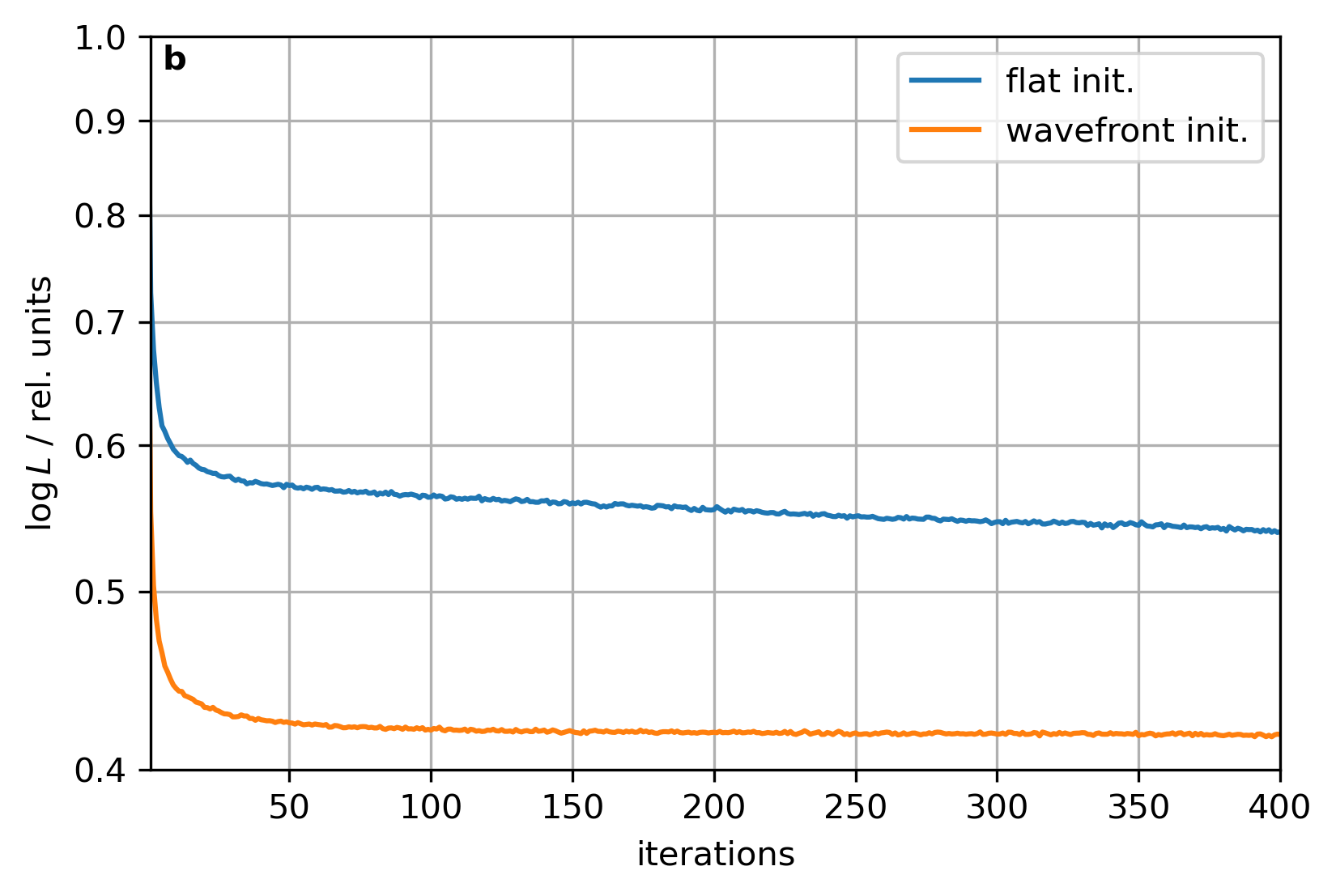}
    \caption{Comparison of the logarithmic value of the cost function $L$ (eq.~\ref{eq:cost_function}) between flat and wavefront initialization as a function of iteration steps for (a) the micrometeorite data and (b) the fluid catalytic cracking catalyst particle. Ptychographic reconstruction with wavefront initialization converges about 200 iteration steps faster in case of (a) and provides superior convergence over the course of 400 iterations in case of (b).}
    \label{fig:micrometeorite_ptycho_error}
\end{figure}

Using wavefront initialization (eq.~\ref{eq:wavefront_init}) implies that the bulk phase information associated with the sample is already present and iteration steps that are associated with the propagation of information from the edges to the center of the sample are skipped during ptychographic reconstruction. Figure~\ref{fig:micrometeorite_ptycho_error} demonstrates that this saves several hundred iteration steps in case of the micrometeorite or even provides superior convergence over the first 400 iteration steps in case of the fluid catalytic cracking catalyst particle. Therefore, ptychographic reconstruction with wavefront initialization can improve iteration speed considerably.

\begin{figure}
    \centering
    \includegraphics[width=0.99\textwidth]{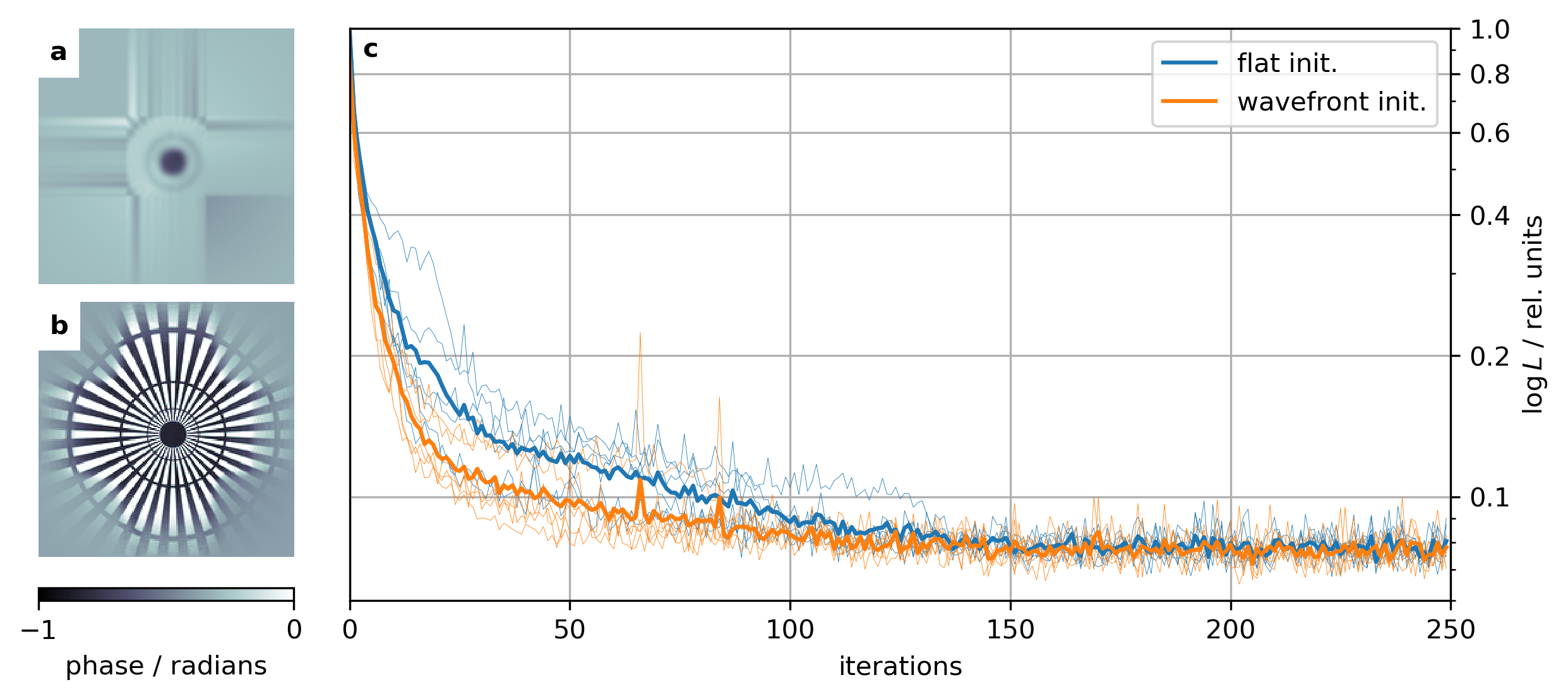}
    \caption{Ptychographic reconstruction of a Siemens star with numerical simulations. (a) Low-resolution phase estimate. (b) Retrieved phase distribution using wavefront initialization. (c) Comparison of the logarithmic value of the cost function $L$ (eq.~\ref{eq:cost_function}) between flat and wavefront initialization as a function of iteration steps. Thick lines are the average values of the cost function of 6 repeated reconstructions (thin lines). While wavefront initialization performs better than flat initialization in the beginning, both take about the same number of iteration steps to converge.}
    \label{fig:siemens_ptycho_error}
\end{figure}

Up to now, the samples included in this study were considerably bulky and, thus, well suited to benefit from ptychographic reconstruction with wavefront initialization. In order to investigate the performance for samples that are more challenging in the present context, we have used numerical simulations of a Siemensstar pattern. In this case, the sample is flat with structures smaller than the probe and, thus, there is no bulk information to propagate during iteration.

The simulated Siemensstar was sampled on a $213\times 213$ grid and provided a minimum transmission of 0.8 and a phase shift of -1.2~rad (Fig.~\ref{fig:siemens_ptycho_error}b), which corresponds to Au structures with a height of 650~nm imaged at a photon energy of 8.2~keV. The probe had a Gaussian-like shape with a full width half maximum of 7~pixels, which is markedly larger than the smallest sample features. The observable diffraction patterns were calculated according to eq.~(\ref{eq:obs_int}) and wavefront retrieval was performed as described above.

Although the estimated wave field for the object used for initialization has insufficient resolution to sample the Siemenstar (Fig.~\ref{fig:siemens_ptycho_error}a), the reconstructed object's phase distribution (Fig.~\ref{fig:siemens_ptycho_error}b) shows that structures smaller than the probe are still reliably retrieved using wavefront initialization. This demonstrates that wavefront initialization is compatible even with challenging samples. However, the comparison of the cost function $L$ between flat and wavefront initialization (Fig.~\ref{fig:siemens_ptycho_error}c) illustrates only a negligible difference in terms of convergence speed. 

\section*{Conclusion}

We have demonstrated that the speed of ptychographic reconstruction algorithms which use a flat initialization for the sample's complex wave field is inherently limited for bulky samples. This is due to the fact that during the iterations the bulk phase information has to propagate from the edges of the sample to its center. By instead using wavefront initialization, the reconstruction speed is considerably increased as the bulk phase information is already present. In addition, we have shown that wavefront initialization can avoid phase singularity artifacts associated with large phase gradients.

The input data for constructing the wavefront initialization is readily accessible in most ptychographic scans via moment analysis of the measured diffraction patterns. The corresponding algorithm for retrieving the complex wave field is -- compared to the ptychographic reconstruction -- fast. In addition, wavefront initialization is readily compatible with a broad range of ptychographic reconstruction algorithms. Taken all together, this makes the presented approach attractive for most ptychography applications.

%This in combination with the obvious statement that all iterative ptychographic algorithms require a starting value for the complex wave field of the sample implies that our approach is readily compatible with a broad range of established ptychographic reconstruction algorithms, which implies a wide spread use.

\section*{Acknowledgements}
We acknowledge DESY (Hamburg, Germany), a member of the Helmholtz Association HGF, for the provision of experimental facilities. Parts of this research was carried out at the PETRA III beamline P06. We thank Dennis Br\"uckner, Stijen van Malderen and Jan Garrevoet for assistance in using P06. This research was supported in part through the Maxwell computational resources operated at Deutsches Elektronen-Synchrotron DESY, Hamburg, Germany.

%\section*{TO (MAYBE) DO}
%\begin{itemize}
%    \item FIGURE: Show result and iteration steps for siemensstar sample, which will %show no speed up but also not drawback -> can be universally used
%    \item cSAXS daten mit hot phase of probe start [2000 Beugungsbilder + refPIE]
%    \item show figure with phase gradients to show that largest lead to phase %singularities as stated in \cite{Wittwer2022}
%    \item use downscaled ePIE as comparison
%    \item SPECTRAL INITIALISATION (YOUTUBE)
%    \item should we not also show a wavefront init reconstruction that acutally uses %less iterations?
%    \item line profile figure with wavefront start?
%    \item put the implementation of wavefront retrieval on github?
%    \item multibeam ptycho~\cite{Bevis2018,Hirose2020}
%\end{itemize}

%\bibliographystyle{unsrt}
\bibliography{references.bib}% Produces the bibliography via BibTeX.

\end{document}